# A Coalition Formation Algorithm for Multi-Robot Task Allocation in Large-Scale Natural Disasters

Carla Mouradian[¥], Jagruti Sahoo[*], Roch H. Glitho[¥], Monique J. Morrow[€], and Paul A. Polakos[£]

[¥]Concordia University, Montreal, Canada,
[*]South Carolina State University, Orangeburg, South Carolina, USA
[€] CISCO Systems, Zurich, Switzerland,
[£]CISCO Systems, New Jersey, USA
ca_moura@encs.concordia.ca, jagrutiss@gmail.com, glitho@encs.concordia.ca {mmorrow, ppolakos}@cisco.com

*Abstract*—In large-scale natural disasters, humans are likely to fail when they attempt to reach high-risk sites or act in search and rescue operations. Robots, however, outdo their counterparts in surviving the hazards and handling the search and rescue missions due to their multiple and diverse sensing and actuation capabilities. The dynamic formation of optimal coalition of these heterogeneous robots for cost efficiency is very challenging and research in the area is gaining more and more attention. In this paper, we propose a novel heuristic. Since the population of robots in large-scale disaster settings is very large, we rely on Quantum Multi-Objective Particle Swarm Optimization (QMOPSO). The problem is modeled as a multi-objective optimization problem. Simulations with different test cases and metrics, and comparison with other algorithms such as NSGA-II and SPEA-II are carried out. The experimental results show that the proposed algorithm outperforms the existing algorithms not only in terms of convergence but also in terms of diversity and processing time.

I. INTRODUCTION

In large-scale disasters scenarios, the primary goal is to find and rescue victims as quickly as possible. Conventional methods involving human rescuers and dogs have well-known limitations. The rescue teams often fail to reach the sites because of a fire break out, collapsed buildings, and closed roads when disaster occur. In contrast, the use of robots brings several advantages. Robots can, for instance, move quickly and find victims more accurately than their human counterparts. The use of robots for search and rescue mission was first witnessed during the rescue operations at the World Trade Center in New York City on September 2001, with the use of CRASAR (Center for Robot-Assisted Search And Rescue) rescue robots [1]. Robots have different capabilities. Forming dynamically an optimal coalition of robots with a required set of capabilities remains very challenging. It is even more challenging in large-scale disaster scenarios. This is due to the very large number of robots [2] needed in a single coalition in order to cover the whole disaster area and satisfy the search and rescue task requirements. The problem can be modeled as a multi-objective optimization problem where the time, the cost, and the number of robots in a coalition are minimized iteratively and simultaneously. In this paper, we propose a coalition formation algorithm to solve it.

We address the ST-MR-IA (Single-Task Multi-Robots Instantaneous-Assignment) class of Multi-Robot Task Allocation (MRTA) problems following the taxonomy presented in [3]. In the problem at hand, each robot is capable of executing one task at a time and each task needs to be assigned to a robot coalition. Also, the available information about the robots, the tasks, and the environment permits only instantaneous allocation of tasks to robots, without planning for future allocations [3]. Our proposed algorithm is based on Quantum Multi-Objective Particle Swarm Optimization (QMOPSO). QPSO is a discrete version of PSO to solve optimization problems with binary-valued solution elements [4]. PSO is one of the many options for coalition formation. Simulated Annealing (SA), Genetic Algorithm (GA), and Column-Generation (CG) are other examples. PSO is chosen because of its effectiveness in solving a wide range of applications [5]. It has the ability to find optimal or near-optimal solutions for large-space problems in a short time compared to other heuristics [6].

The goal of the proposed algorithm is to ensure that the optimal coalition of robots is selected with the required capabilities for each task. The proposed algorithm consists of filtering method, the QMOPSO approach, and ranking method. Filtering is used to choose the best robots for the execution of the QMOPSO algorithm and to make the robots that have not been selected available for other requests. In addition, location constraints regarding the capability distribution of the robots are taken into consideration. For instance, some tasks require that the combination of a given sensor and actuator should reside on the same robot or on different robots. This is necessary to ensure proper execution of search and rescue task. The rest of the paper is organized as follows: Section II discusses related work. In section III, the problem is formulated. Section IV presents the proposed algorithm. Validation is done in Section V and we conclude in Section VI.

II. RELATED WORK

Forming an effective coalition has become an interesting research area. It has been widely studied over years. The problem is NP-hard [7]. Reference [8] proposes a modified version of Shehory and Kraus's algorithm. Authors in [9], propose an algorithm based on GA. Their major drawback is that they optimize only one objective (i.e., the overall utility and the coalition value respectively). However, there are other important objectives that need to be optimized, e.g., the time needed to perform a task. Reference [10] proposes an algorithm which aims at maximizing the number of completed tasks and the system efficiency. Two multi-objective algorithms are introduced: a Non-Dominated Sorting Genetic Algorithm (NSGA-II) and a Strength Pareto Evolutionary Algorithm (SPEA-II). Unfortunately, factors such as the minimizing the number of robots in a coalition is not considered. In [11], authors assign Unmanned Aerial Vehicles (UAVs) to search and prosecute missions. However, reducing the cost of UAV deployment is not considered. Authors in [12] propose PSO-based multiple objectives algorithm. The PSO is designed for continuous domains. We consider a discrete problem domain for the robots, which drives the need for designing new PSO. In [13], authors propose an auction-based approach for task allocation. However, their algorithm is evaluated over small scale i.e., 10 and 50 robots. Additionally, none of the works presented above discusses any filtering or ranking methods. In our work, we present ranking and filtering methods for the best solution among Pareto-optimal solutions, to optimize the processing time of the algorithm. In [14], authors propose an ant-colony based algorithm. They consider fix number of robots for each task. However, it is not efficient to fix the number of robots required for each task since robots have different capabilities and different capability distribution. Authors in [15], propose an algorithm based on dynamic ANT coalition technique. However, the performance of the algorithm is not evaluated with a large number of robots. In [16], Rauniyar and Muhuri modify the standard GA. They proposed adaptive Random Immigrants Genetic Algorithm (aRIGA) and adaptive Elitism-based Immigrants Genetic Algorithm (aEIGA). However, the proposed algorithm is single-objective.

III. PROBLEM FORMULATION

We consider an infrastructure composed of *n* robots:
$$R = \{R_1, \dots, R_i, \dots, R_n\} \quad (1)$$
where *n* is significantly large and hence the infrastructure can support search and rescue task in large-scale disasters.

Each of these robots has two vectors of capabilities: sensing (e.g., cameras, sensors) and actuating (e.g., arms, wheels) capabilities. It is assumed that each capability is a real non-negative value and indicates the quantity of sensors/actuators owned by the robots.

For robot $R_i$, the sensing and actuating capability vectors are:
$$S_{R_i} = \{s_1^i, \dots, s_r^i\} \quad (2) \qquad A_{R_i} = \{a_1^i, \dots, a_d^i\} \quad (3)$$
where *r* and *d* are the number of possible sensing and actuating capabilities respectively.

A robot can be in three states: *Idle*, *Allocated*, and *Busy*. The idle state is when the robot does not perform any tasks, the allocated is when the robot is locked with the algorithm running on it, and the busy state is when the robot is performing a task.

The infrastructure can perform *m* tasks assigned to it:
$$T = \{T_1, \dots, T_j, \dots, T_m\} \quad (4)$$

Each task $T_j$ is composed of *p* sub-tasks:
$$Z_{T_j} = \{z_1^j, \dots, z_k^j, \dots, z_p^j\} \quad (5)$$

It is assumed that the sub-tasks are executed independently and that each robot is a member of only one sub-task. Each sub-task requires a specific set of sensing and/or a set of actuating capabilities to start.

We represent the capability requirements of each sub-task $z_k^j$ by two vectors, sensing requirements and actuating requirements, as:

For sub-task $z_k^j$, the capability requirement vectors are:
$$S_{z_k}^j = \{s_1^{jk}, \dots, s_r^{jk}\} \quad (6) \qquad A_{z_k}^j = \{a_1^{jk}, \dots, a_d^{jk}\} \quad (7)$$

Then, the capability requirement vectors for the task $T_j$ is the sum of the capability requirement vectors of the sub-tasks constituting the task $T_j$:
$$S_{T_j} = \sum_{k=1}^p S_{z_k}^j \quad (8) \qquad A_{T_j} = \sum_{k=1}^p A_{z_k}^j \quad (9)$$

Some of the sub-tasks of task $T_j$ are tied by locational constraints regarding the capability distribution of the robots while others may be executed without any locational constraints. This is necessary in order to ensure proper execution of the sub-tasks. According to [8], there are two types of locational constraints; combination of sensors and actuators should reside on the same robot or combination of sensors and actuators should reside on different robots. The locational constraints can be represented as Constraints Satisfaction Problem (CSP) [8].

CSP consists of three components:
- The set of variables, that is the required sensor and actuators for the task
$$X = \{x_1, \dots, x_j, \dots, x_k\} \quad (10)$$
where $X = \{s_1, s_2, \dots, s_n, a_1, a_2, \dots, a_n\}$
- The set of values for each variable, that is the available robots possessing the required capabilities for each variable

For variable $x_i$, the set of values is:
$$V_{x_i} = \{R_j, \dots, R_n\} \quad (11)$$
- The set of constraints between different variables
$$C = \{C_1, \dots C_i, \dots, C_n\} \quad (12)$$
where each $C_i$ is one of the following types: $x_i \neq x_j, x_i = x_j$. The goal in CSP is to assign a value for each variable such that the constraints are satisfied.

A coalition $m$ ($CLT^m$) for any task has two vectors of capabilities: sensing $S_{CLT^m}$ and actuating $A_{CLT^m}$, while each is the sum of the capabilities owned by the robots in that coalition:
$$S_{CLT^m} = \sum_{R_i \in CLT^m} S_{R_i} \quad (13) \qquad A_{CLT^m} = \sum_{R_i \in CLT^m} A_{R_i} \quad (14)$$

$CLT^m$ can perform task $T_j$ only if:
1. The vector of its capabilities satisfy the following:
$$S_{CLT^m} \geq S_{T_j} \quad \text{And/or} \quad A_{CLT^m} \geq A_{T_j} \quad (15)$$
2. And its members meet the locational constraints.

It is assumed that a coalition can work on a single task at a time and that each robot ($R_i$) is a member of one coalition at a time. Where $i = [1, n]$.
$$CLT_{R_i}^1 \cap CLT_{R_i}^2 = \emptyset \quad (16)$$

The objective of this problem is to find a coalition of robots that minimizes simultaneously the deployment cost of the robots, the time needed to perform a task by the robots belonging to that coalition, and the number of the robots in a coalition in the same amount of time.

## IV. THE PROPOSED ALGORITHM FOR COALITION FORMATION

We propose an algorithm for coalition formation problem. Our optimization problem includes constraints, including the task capability requirements and locational constraints as described in section III. A particle is said to be feasible if it satisfies both constraints and infeasible otherwise. Given a particle (i.e., candidate coalition), checking the task capability requirements is quite straightforward. However, the locational constraints are checked by solving CSP, which can be solved with either the brute-force technique or a better technique such as backtracking. Handling constraints during QMOPSO execution is described in the next section (section IV. B).

### A. Coalition Formation Algorithm for Multi-Robot System

The Pseudocode for the coalition formation algorithm for multi-robot systems is given in Algorithm 1. The detailed description is as follows: The set of inputs for the algorithm are n (the maximum number of robots allowed in a group), Time (the maximum time period to complete a given task), Cost (the cost the customer agrees on), Filtering_Rule, Task_Requirements (the required sensors and actuators for a given task), Locational Constraints (the capability distributions for sub-tasks constituting a given task), and Criteria_Importance (defining the weights to rank the Pareto-optimal solutions based on more than one criterion - i.e., objectives in our case).

The algorithm starts with filtering the robots based on the Filtering_Rule. In this function, if the battery level of the robots is lower than the Filtering_Rule, they are excluded from the next steps. It then applies the QMOPSO-based algorithm. Multi-objective problems generate a set of non-dominated or Pareto-optimal solutions. The solutions are ranked after excluding the solutions that exceed the time, the cost, the number threshold, and the infeasible solutions. Promethee II ranking [17] is applied, which is a multi-criteria ranking method with lots of success due to its mathematical properties and its user-friendliness. In this method, the Pareto-optimal solutions are compared pairwise. The difference between the evaluations of two Pareto-optimal solutions over each criterion is considered. The criteria in our case are the objectives (i.e., time, cost, and number of robots). The Pareto-optimal solutions are ranked using the Criteria-Importance/weight of the objectives. The highest rank solution denotes the best robot coalition.

### B. QMOPSO Algorithm

The Pseudocode of QMOPSO is given in Algorithm 2. The algorithm first initializes the particles. A particle is defined based on the quantum bit. Two vectors are initialized:

- Quantum particle vector $V(t)^i$, which is the velocity for particle $i$ and is initialized to random values between [0,1]:
$$V(t)^i = [v(t)^i_1, \ldots, v(t)^i_j, \ldots, v(t)^i_n] \quad (17)$$
- Discrete particle vector $p(t)^i$, which is initialized by initializing

**Algorithm 1: Coalition Formation Algorithm for Multi-Robot Systems**
1. Inputs: n, time, cost, Filtering_Rule, Task_Requirements, Criteria_Importance, allRobots
2. Set Selected_Robots = [ ], Selected_Clts=[ ], Robots=[ ]
3. **Function**: Filtered_Robots = Filter_Robots (allRobots, Filtering_Rule)
4. Selected_Robots =Filtered_Robots
5. **Function**: apply QMOPSO to find the best coalition
6. **foreach** (Robot in best coalition)
7.     set Robot.State = busy
8.     deploy Robot
9. **end for**
10. **if** more than one Pareto-Optimal solution **then**
11.     **if** time, cost, number of robots for each particle exceed thresholds (t, c, n) → remove particle
12.     **else if** particle is infeasible → remove particle
13.     **else**
14.         Rank the Pareto-Optimal solutions
15.         Select particle with highest ranking
16.         selected_coalition = Particle with highest ranking
17.     **end if**
18. **else if** one Pareto-Optimal solution
19.     selected_coalition = the Pareto-Optimal solution
20. **end if**

**Algorithm 2: QMOPSO-based Heuristic Algorithm**
1. Initialize number of iteration, $V(t)$ and $P(t)$
2. t=0
3. value = Evaluate Population($P(t)$)
4. Store the position of particles that represents non-dominated vector in repository REP
5. Initialize memory for each particle
6. $p_{i_{localBest}}[i] = P_i(t)$
7. $t = t + 1$
8. **while** (number of iteration is not reached) **do**
9.     Set $P_{globalBest}$ by selecting from the REP
10.    **foreach** particle$P(t)$
11.        Update velocity and position of particles
12.        value = Evaluate Population
13.        Update the$P_{localBest}$
14.        **if** the current $P(t)$is non-dominated by $p_{i_{localBest}}[i]$
15.            $p_{i_{localBest}}[i] = P_i(t)$
16.        **end if**
17.    **end for**
18.    select the non-dominated particles
19.    Update the REP by comparing current non-dominated particles with the ones in REP
20. **end while**

a random number for each $v(t)^i_j$ and then, according to the condition in (19) and (20), the discrete particle vector is initialized.
$$p(t)^i = [p(t)^i_1, \ldots, p(t)^i_j, \ldots, p(t)^i_n] \quad (18)$$
where *n* is the size of the problem, i.e., the total number of robots.

$$\text{If} \quad rand^i_j > v(t)^i_j \rightarrow p(t)^i_j = 1 \quad (19)$$
$$\text{Otherwise} \quad p(t)^i_j = 0 \quad (20)$$

First, the initial population is evaluated by calculating the values of three objective functions for each particle. The particles that represent non-dominated solutions are stored in a repository (*REP*). Each particle keeps track of its best local position, which is the best solution obtained by this particle so far ($P_{i_{localBest}}$). At each iteration, the algorithm selects $P_{globalBest}$ that denotes the best position achieved so far by any particle in the population. It is selected by ranking the solutions in REP and choosing the one with the highest rank. Also, the velocity equation is updated according to equation (21) and the particle vector is updated in the same way in equations (17) to (20).

$$V(t+1) = w \times V(t) + c_1 \times V_{localbest}(t) + c_2 \times V_{globalbest}(t) \quad (21)$$

TABLE I. ALGORITHM EVALUATION PARAMETERS

| Parameter | Value |
|---|---|
| **General** | |
| Population size | 100, 200 |
| Problem size (number of robots) | 10-10000 |
| Maximum number of iterations | 100 |
| $\alpha, \beta$ | 0.3, 0.7 |
| $w, c_1, c_2$ | 0.25, 0.25, 0.5 |
| Threshold for filtering | 40% |
| Number of objectives | 3 |
| Number of sub-tasks | 3 |
| Criteria_Importance | time |
| **NSGA-II and SPEA-II** | |
| Tournament size | 2 |
| Pool size for tournament selection | Population number / 2 |
| Mutation probability | 10% |
| Crossover probability | 90% |
| Distribution index for crossover | 20 |
| Distribution index for mutation | 20 |

TABLE II. ERROR RATIO, SET COVERAGE AND SPACING (10 ROBOTS, POPULATION SIZE=200)

| Algorithm | Error Rate | Set Coverage | | | Spacing |
|---|---|---|---|---|---|
| | | QMOPSO | NSGA-II | SPEA-II | |
| QMOPSO | 0.6 | - | 0.3 | 0 | 31.63 |
| NSGA-II | 0.8 | 0.66 | - | - | 51.45 |
| SPEA-II | 0.33 | 1 | - | - | 67.14 |

TABLE III. ERROR RATE & SET COVERAGE (POPULATION SIZE=100)

| No. of Robots | Error Ratio | | | Set Coverage | | | |
|---|---|---|---|---|---|---|---|
| | QMOPSO | NSGA-II | SPEA-II | (QMOPSO, NSGA-II) | (NSGA-II, QMOPSO) | (QMOPSO, SPEA-II) | (SPEA-II, QMOPSO) |
| 1000 | 0.23 | 0.32 | 0.30 | 0.72 | 0 | 0.8 | 0.1 |
| 5000 | 0.20 | 0.31 | 0.41 | 0.9 | 0.1 | 1 | 0 |
| 10000 | 0.11 | 0.26 | 0.2 | 1 | 0 | 0.93 | 0 |

TABLE IV. SPACING (POPULATION SIZE=100)

| No. of Robots | Spacing | | |
|---|---|---|---|
| | QMOPSO | NSGA-II | SPEA-II |
| 1000 | 18.23 | 23.11 | 40.36 |
| 5000 | 16.11 | 35.61 | 37.22 |
| 10000 | 8.24 | 19.23 | 28.19 |

$$V_{localbest}(t) = \alpha \times p_{localbest}(t) + \beta \times (1 - p_{localbest}(t)) \quad (22)$$

$$V_{globalbest}(t) = \alpha \times p_{globalbest}(t) + \beta \times (1 - p_{globalbest}(t)) \quad (23)$$

where $\alpha + \beta = 1$, $\beta < 1$, $0 < \alpha$. $\alpha$ and $\beta$ are control parameters, $w$ represents the degree of belief on oneself, $c_1$ is the local maximum, and $c_2$ is the global maximum. The first part of the equation (21) indicates the interia of the previous probability. The second part is called cognition and represents the local exploration probability. The third part is the social part that indicates the cooperation among all quantum particles.

$P_{i_{localbest}}$ is updated by applying Pareto dominance. If the current position is dominated by the one in the memory, the one in the memory is kept; otherwise, the one in the memory is replaced by the current position.

To update the *REP*, if the *REP* is empty, the current non-dominated particle is added to the *REP*; otherwise, the two particles are compared as follows: If both are feasible, Pareto-dominancy is applied; if one is feasible and the other infeasible, the feasible dominates; if both are infeasible, the one with the highest degree of constraint satisfactions is selected. We define a particle's feasibility degree as the degree of constraint satisfactions. A task $T_j$ is considered to have *U* capability requirements and *M* locational constraints. The capability requirements and locational constraints are considered as $T_{req}$ and *C* respectively. Then a particle is feasible if it satisfies $T_{req}$ and *C*, and it is infeasible otherwise. We determine a particle's feasibility degree as the weighted sum of feasibility degree with respect to $T_{req}$ and *C*.

If a particle satisfies *u* capability requirements and satisfies *m* locational constraints, then the particle's feasibility degree with respect to $T_{req}$ and *C* are expressed as:

$$sat_{T_{req}} = u/U \quad (24) \qquad sat_C = m/M \quad (25)$$

A particle's feasibility degree can now be calculated as:

$$P_{feas} = sat_{T_{req}} * W_T + sat_C * W_C \quad (26)$$

where $W_T$ and $W_C$ are the weights chosen such that:

$$W_T + W_C = 1, \quad 0 \leq W_T, W_C \leq 1 \quad (27)$$

If a particle is feasible, then the feasibility degree is 1.

## V. PERFORMANCE EVALUATION

In order to evaluate the algorithm, we have performed our experiments with different problem and population sizes. In each experiment, the speed, the cost, the position, the battery level of each robot, and the position of the target - which is the disaster location – have been randomly generated. All the robots are in the idle state at the beginning of each experiment. We have compared our algorithm with two well-known heuristic-based algorithms: NSGA-II and SPEA-II [10]. All algorithms have been implemented in Matlab. Table I shows the evaluation parameters along with their values.

### A. Performance Metrics

We measured the performance of three optimization algorithms in terms of convergence, diversity, and processing time. Convergence shows the solution's accuracy (i.e., the ability to produce good quality solutions and their ability to converge to true Pareto-solutions). We have used error ratio [12] and set coverage as the convergence metrics. Diversity, on the other hand, shows the spread of solutions. We have used spacing [12] as the diversity metric. The metrics are defined as follows:

1. *Error Ratio,* as the percentage of non-dominated solutions that are not part of a reference Pareto-set: When the true Pareto-set is known, it is used as the reference Pareto-set. When it is not known, the reference Pareto-set is obtained by combining Pareto-sets of all algorithms and applying non-dominancy.
2. *Set Coverage (SC (A, B)),* given two sets, is the percentage of non-dominated solutions in set B covered (i.e., dominated) by those in set A: If SC (A, B) > SC (B, A), then A is relatively better than B. A is absolutely better than B when SC (A, B) = 1 and SC (B, A) = 0.
3. *Spacing,* as the standard deviation of distances of non-

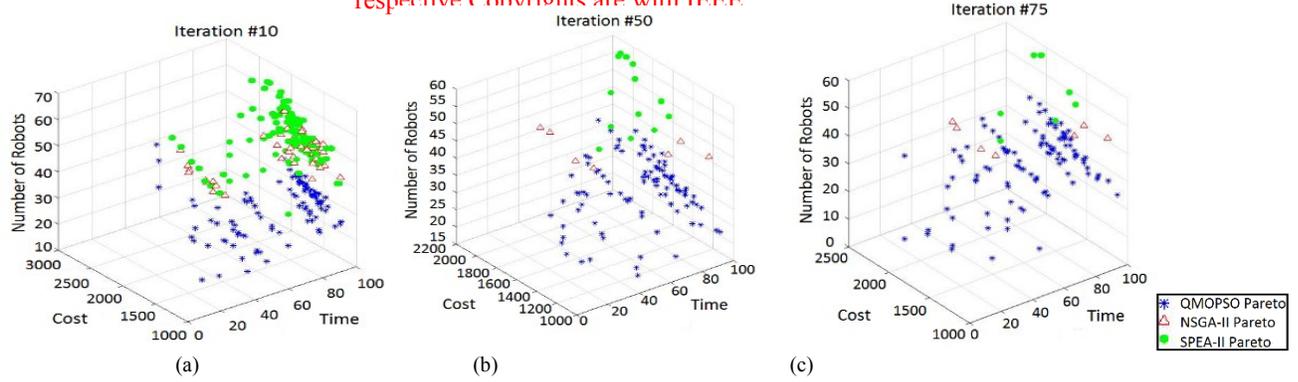

Fig. 1. Non-dominated fronts obtained at different iteration for problem size 5000 and population size 200

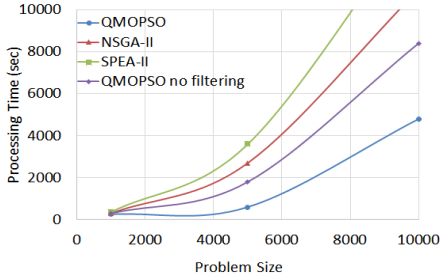

Fig. 2. Processing time with different problem sizes (Population size=100)

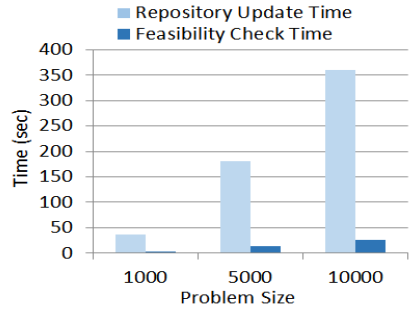

Fig. 3. The effect of feasibility check on average repository updating time

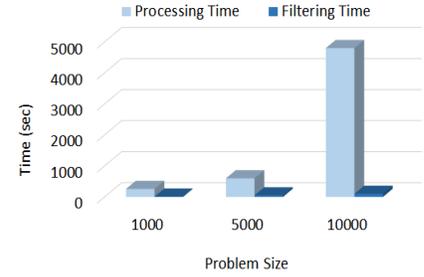

Fig. 4. The time needed for the filtering function in QMOPSO

dominated solutions from their closest neighbors.
4. *Processing Time of Algorithm (PT)* (Sec), as the time needed for the algorithm to run and select the most efficient coalition.
5. *Filtering Time* (sec), as the time needed for filtering the robots in QMOPSO.
6. *Repository Update Time* (sec), as the time needed by QMOPSO to update the repository at each iteration. It includes the delay incurred by the constraint handling method.

B. Results and Discussions

   *Test case 1 - convergence and diversity:* Table II shows the error ratio, the set coverage, and the spacing metrics of QMOPSO, NSGA-II, and SPEA-II for a small-scale problem (i.e.., 10 robots). We have also generated the true Pareto-optimal solutions by the enumerated search method. For the three algorithms, we have used a population size of 200. The error ratio of QMOPSO is higher than that of SPEA-II but lower than that of NSGA-II. The set coverage metric shows that 30% of solutions in NSGA-II are covered by QMOPSO and 66% of QMOPSO are covered by NSGA-II. Hence, NSGA-II is relatively better than QMOPSO. Since NSGA-II and SPEA-II do not cover each other, their relative dominance cannot be concluded. Overall, we observe that NSGA-II performs better than QMOPSO in terms of convergence. However, when it comes to diversity, QMOPSO outperforms NSGA-II and SPEA-II. This is concluded from the lowest spacing value in case of QMOPSO, which indicates a good distribution of solutions. Table III shows the error ratio for large-size problems (e.g., 1000, 5000, and 10000 robots). We observe that for any problem size, QMOPSO outperforms NSGA-II and SPEA-II. In fact, it achieves the lowest error ratio for the largest problem size (10000 robots). It shows a better convergence of QMOPSO for large-scale problems. SPEA-II has the highest error ratio for 5000 robots. Table III also shows the set coverage metric. As observed, when the problem size is 1000, QMOPSO is relatively better than both NSGA-II and SPEA-II. For a problem size of 5000, QMOPSO is absolutely better than SPEA-II as all solutions of SPEA-II are dominated by those of QMOPSO and none of QMOPSO solutions is dominated by those in SPEA-II. QMOPSO for a problem size of 5000 is relatively better than NSGA-II. However, for a problem size of 10000, QMOPSO is relatively better and absolutely better than SPEA-II and NSGA-II respectively. Overall, SC results show that QMOPSO produces a better solution than NSGA-II and SPEA-II. Table IV shows the spacing metric for three algorithms. We observe that QMOPSO attains the lowest value of spacing for any problem size, thereby achieving the highest diversity and even distribution of solutions. The diversity of NSGA-II lies between QMOPSO and SPEA-II. Fig. 1 (a)-(c) shows the non-dominated-fronts obtained at some iterations for a problem size of 5000. We have found that with an iteration increase, the solutions in QMOPSO evolve more quickly than those in NSGA-II and SPEA-II. It shows the ability of QMOPSO to explore the search space more efficiently than others.

   *Test case 2 - processing time of the algorithms with a various number of robots:* We compare the PT of our algorithm with NSGA-II and SPEA-II. The three algorithms are implemented and applied in the same environment, with the same number of robots, task requirements, and robot capabilities. The size of the population is 100. Fig. 2 shows the processing time of the three algorithms with a various number of robots. For the QMOPSO algorithm, we consider the PT with and without the filtering method. We notice that the PT decreases when the filtering method is used. This is because filtering method reduces the number of robots on which the algorithm runs. On the other hand, without filtering, the PT of the algorithm increases with an increase in the number of robots. The rationale behind this is the fact that the higher number of robots results in a higher dimension of the particle. As an important observation, the PT without these methods is still smaller than that of NSGA-II and SPEA-II; this is due to the simple mathematical operations of QMOPSO compared to other algorithms. In QMOPSO, the velocity equation is the sole equation updated at each iteration.

*Test case 3 - repository update time:* We have considered two types of constraints: Task requirements and location constraints. For the task requirements, we have considered 6 requirements ($s_1, s_2, s_3, a_1, a_2, a_3$) with random number of units for each. For the locational constraints, we have represented the problem using CSP as described in section III and we have considered three locational constraints ($s_1 = a_1, s_2 = a_2, s_3 = a_3$). A simplified method is used to calculate the satisfaction degree of a particle/coalition for the task requirements and locational constraints. We calculate the effect of our proposed method to solve the two constraints (task requirements and locational constraints) on the average repository updating time. The results in Fig. 3 demonstrate the time needed to perform the feasibility check versus the overall repository update time, considering different numbers in a population. As we notice, the time needed for our proposed feasibility checking method is negligible compared to the total time for updating the repository.

*Test case 4 - filtering time:* We have also calculated the time needed to perform the filtering function compared to the overall processing time of QMOPSO. Fig. 4 shows that the filtering time is negligible compared to the overall processing time of the algorithm. The time needed for filtering does not introduce additional overhead on the algorithm processing time. Since this method excludes some robots using battery levels, it ensures that the remaining robots have sufficient battery to accomplish the task. Since it does not affect the processing time of the algorithm, the overall efficiency is achieved.

## VI. CONCLUSION

In this paper, we have proposed a coalition formation algorithm for multi-robot systems. To show the effectiveness of our algorithm, we have conducted extensive simulation experiments and compared our algorithm with other existing algorithms. The results demonstrate that the proposed algorithm cannot only improve the solution, but it also has a significantly short processing time. They also show that the filtering and the repository updating mechanisms do not add overhead on the processing time. It is also observed that QMOPSO achieves higher diversity, the lowest error rate, and produces better solution compared to NSGA-II and SPEA-II for large problem sizes.

ACKNOWLEDGMENT

The work is partially supported by CISCO systems grant CG-576719 and by the Canadian National Science and Engineering Research Council (NSERC) discovery program.